# ARChef: An iOS-Based Augmented Reality Cooking Assistant Powered by Multimodal Gemini LLM

Rithik Vir[1†], Parsa Madinei[2,3]
*[1]Cupertino High School, 10100 Finch Ave, Cupertino, CA 95014*
*[2]Department of Psychological & Brain Sciences, University of California, Santa Barbara, CA 93106*
*[3]Department of Computer Science, University of California, Santa Barbara, CA 93106*
[†]*corresponding author: rithikvir@gmail.com*

Cooking meals can be difficult, causing many to resort to cookbooks and online recipes. However, relying on these traditional methods of cooking often results in missing ingredients, nutritional hazards, and unsatisfactory meals. Using Augmented Reality (AR) can address these issues; however, current AR cooking applications have poor user interfaces and limited accessibility. This paper proposes a prototype of an iOS application that integrates AR and Computer Vision (CV) into the cooking process. We leverage Google's Gemini Large Language Model (LLM) to identify ingredients in the camera's field of vision and generate recipe choices with detailed nutritional information. Additionally, this application uses Apple's ARKit to create an AR user interface compatible with iOS devices. Users can personalize their meal suggestions by inputting their dietary preferences and rating each meal. The application's effectiveness is evaluated through three rounds of user experience surveys. This application advances the field of accessible cooking assistance technologies, aiming to reduce food wastage and improve the meal planning experience.

**Keywords:** Augmented Reality, Computer Vision, Large Language Models, Google Gemini, User Interface

## I. INTRODUCTION

Despite the importance of cooking meals, using traditional cooking methods can be a struggle for many, especially those with limited cooking experience. In fact, a 2013 study in the United Kingdom identified that 24.4 percent of men, along with 7 percent of women, are not confident in their abilities to cook [1]. Additionally, access to cooking aids such as robot cooking assistants or private chefs are extremely limited due to their cost, leading to nearly half of all 1.03 billion pounds of food waste being generated at the household level [2]. Due to the limited accessibility of cooking assistants, many resort to using conventional cooking methods such as cookbooks and online recipes; however, doing so results in a struggle to find a recipe that can be made using only the ingredients the user has. The result of this inconvenience is unsatisfactory meal choices, food wastage, and the skipping of meals entirely [3]. In addition, nutritional information, including potential allergens and intolerances, isn't always displayed in these conventional methods, which can cause chronic health problems and allergic reactions [4].

By creating an accessible cooking aid that can display meal options using only the ingredients the user has, food wastage and hunger can be mitigated. Using Augmented Reality (AR) technology, can serve as a more accessible solution to this problem, especially when combined with Computer Vision (CV) to identify and display the ingredients at hand. Additionally, Artificial Intelligence (AI) enables users to obtain high-quality meal options along with detailed nutritional information. The result of this combination, which also incorporates principles regarding the user interface from Milgram and Kishino's Reality-Virtuality (RV) continuum, is a hassle-free user interface that provides detailed instructions for cooking specific meals [5].

There have been several studies creating AR user interfaces for cooking, including one that attempted to create a personalized application tailored towards users [6]. In this previous study, Iftene et al. developed an application called AREasyCooking which took in the user's dietary restrictions and favorite cuisines to create a personalized application for each user. However, when recommending recipes, the AREasyCooking application would recommend recipes from a pre-existing database, rather than using AI to generate them, not addressing the problem of users being able to find a recipe that uses all of their ingredients. Another study attempted to build off of this by using the Magic Leap One AR headset and the YOLOv5 deep learning model to scan ingredients and recommend recipes [7]. However, the accuracy of the YOLOv5 model when detecting ingredients was relatively low, as it was prioritized for speed, so it had difficulty when detecting certain ingredients. Additionally, the user would have to purchase the $550 Magic Leap One headset in order to access the AR cooking assistant which doesn't help with the fact that cooking assistants are already expensive and inaccessible themselves.

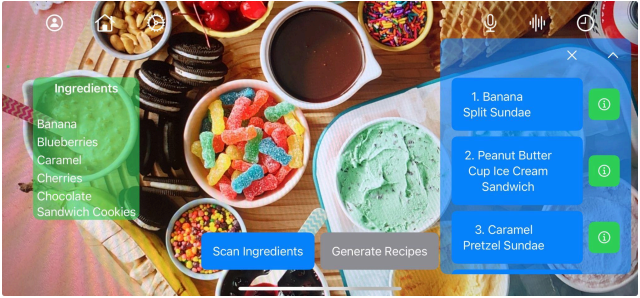

FIG. 1. Layout of the application's components: scanned ingredients and recipe suggestions

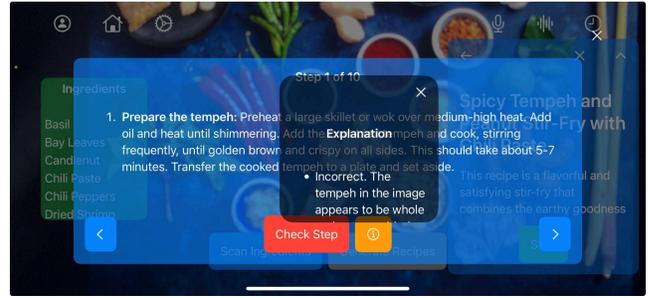

FIG. 2. Example feedback on a recipe step done incorrectly

In this paper, we create a prototype of an AR cooking application that can be accessed simply through an iOS device. We developed the application to detect the user's ingredients and generate detailed recipes that can be made using only what the user has. The application also can be personalized to the users through the settings page, where users can select their language and nutritional preferences, such as allergies and food intolerances. The recipes displayed by the application always contain nutritional information, including the amount of calories, fat, carbohydrates, and protein. Users can also ask follow-up questions using our voice and text-based assistants. The application also features interactive cooking aids such as timers and shopping list generators for when the user already has a recipe in mind.

Our work is the first to combine iOS-compatible AR, Google's Gemini LLM, and personalized nutrition tracking in a single cooking assistant application. We aim to create an accessible cooking assistant to reduce the struggle of cooking, making audiences more eager and knowledgeable when cooking meals.

## II. METHODS

To develop our application, we used Apple's ARKit programming interface via Xcode 15.4. We used the Google AI Software Development Kit to access the Google Flash 1.5 Gemini model in our application and created the API key through Google AI Studio.

For the first component of our application, the live-AR labeling for ingredient detection, we utilized the ability of the Gemini model to process multimodal inputs and attached a snapshot of the device's screen along with a prompt asking for the model to return the coordinates of each ingredient. Using the coordinates received, we added labels to the screen above each ingredient and added the ingredients detected to a list, which contained the total list of ingredients identified over time.

We used this list to attach to the prompt when prompting the model to return the list of optimal recipes in order to optimize accuracy when suggesting meals. The Gemini model is also prompted to return a detailed recipe for each meal suggestion that clearly outlines the ingredients and amounts needed. An example of the layout of the scanned ingredients and the suggested recipes is shown in Figure 1.

In addition, our application also features an AI assistant available through the forms of a voice assistant and a chatbot. Both assistants stem from the same Gemini LLM and also use system instructions to tailor the behavior of each model. These assistants use the same system instructions that form them to serve as the role of a knowledgeable sous chef who provides cooking advice, recipe suggestions, and answers food-related questions. Both these assistants have access to the ingredients the user has scanned, the recipes they have been recommended, and the selected recipe.

To optimize the user experience from the proposed application, the Gemini model calculates the nutritional information that includes the amount of calories, fat, carbohydrates, fiber, protein, and various vitamins. The Gemini LLM then displays this information along with possible allergens next to each recipe in order to prioritize user safety.

In order to help navigate users throughout the recipes, our application includes the feature to check if a user has done a step in the recipe correctly. This is done by having the model receive a snapshot of the user's workspace, along with a prompt for it to provide feedback on the user's performance. An example of the feedback provided is displayed in Figure 2.

Our application is designed to be used either as a handheld tool by having users interact with the app by using buttons to capture snapshots of their environments while interacting with the heads-up displays located around the screen. Users can also have back-to-back conversations with the AI voice assistant to generate recipes and ask follow-up questions on each recipe with the press of a button, to prompt it to listen to what the user is saying.

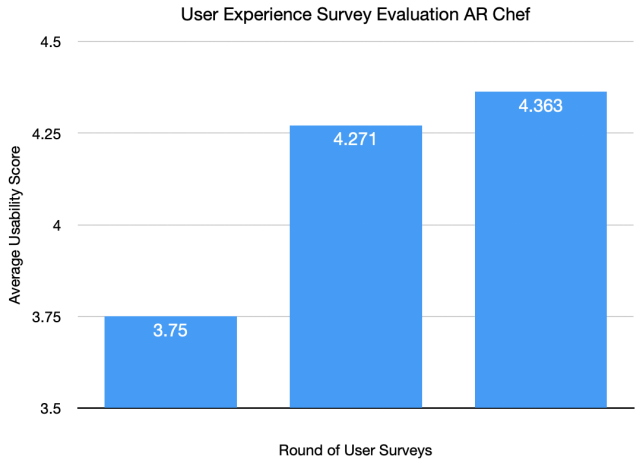

FIG. 3. Survey results for each round of user experience testing

The application also includes a page where participants can input their dietary restrictions, favorite cuisines, and cooking level to tailor the recipes recommended to them.

Both the voice and text assistants, along with the recipe generation model have access to the information in this page and can alter their suggestions based on it.

To increase the reach of our application to a wide audience, it is accessible in eight different languages: English, Spanish, French, Chinese, Japanese, Arabic, Persian, and Hindi. This was done by setting hard values for each button and setting while using the Gemini model to translate each of the responses it generated. We believe this feature will help users better understand the recipes for their ethnic cuisines.

In order to continuously optimize the user experience for our application, we conducted a series of surveys to volunteers in order to receive feedback using the 5-point Likert scale. With this survey, we received reports from 21 total participants through 3 rounds of surveying 6, 7, and 8 participants each. Our participants ranged from high school and college students to working adults and senior citizens.

The surveys used to evaluate the user experience were adapted from the study conducted by Majil et. al [7], and each contained two sets of questions. The first set of questions evaluated the cooking background of each participant in order to determine the diversity of the participants, while the second set of questions asked questions regarding the user experience of the application and offered a space for participants to enter feedback. After each round, we implemented the feedback we received through the surveys in our application and compared the results from each round of surveys. This was done by taking the sum of each question's score on the 5-point Likert scale and taking the average for the participants in each round, resulting in a score out of 5.

## III. RESULTS & DISCUSSION

In this paper, we surveyed a total of 21 volunteers, ranging from high schoolers to senior citizens. These volunteers rated their knowledgeability and frequency of cooking to be an average of 2.714 and 2.857 on the 5-point Likert scale, respectively. Additionally, the participants noted that they often struggle when deciding what to cook, as they chose an average score of 4.524 out of 5 when asked this question.

We evaluated the user experience of the volunteers by taking their average score on each of the questions in the second section of the survey, which evaluated their experience using the application. In the first round of surveys, the 6 participants had an average usability score of 3.75 out of 5, demonstrated in Figure 3, indicating that our application somewhat helped their cooking experience. These participants also provided feedback on the application and what could be improved. The advice we implemented into the next round of surveys was that they "wished the windows self-organized themselves so [they're] easy and clear to see." Another piece of feedback we implemented into the next version of the application was that we included "a way to remove items from the [ingredient list] or manually add an ingredient that was misclassified."

In the second round of user experience surveys, Figure 3 displays that the average usability score of the next 7 participants increased to 4.27 out of 5. The factors that affected this increase in the usability score the most were how easy it was to learn how to use the application, as well as whether the participants were able to carry out system functions without difficulties or errors. The primary factor for these improvements was the organization that was implemented in between these surveys, as we gave the application a more rigid structure by organizing the windows for the user rather than leaving it up to them. Additionally, we conducted a round of bug-fixing between these rounds of surveys, which reduced the amount of errors and inconveniences in the application.

By the third round of interviewing the participants, we implemented the personalization feature of our application where recipes would be tailored to users dietary restrictions, favorite cuisines, and cooking level. This resulted in an average usability score of 4.36 out of 5, indicating that the average user was very satisfied with our application. Furthermore, every user in this round reported that they were satisfied with our application and would recommend it to others.

Overall, throughout each section of the surveys, each participant was at least somewhat satisfied with our application, and by the last round, every participant either agreed or strongly agreed that they would use our application again.

By utilizing our application, users from all levels of cooking experience can eliminate the hassle of finding a recipe using only the ingredients they have by using the feature of our app to scan their ingredients and generate recipes for them. Users can also use our application as an accessible cooking assistant as it features both text and voice assistants that know which ingredients they have along with the recipes they've been displayed and chosen. Additionally, our application provides cooking aids such as timers and also contains a feature to evaluate and explain if the user did each step correctly through the click of a button. Even if the users already have a recipe in mind, the application can generate shopping lists for the extra ingredients the users need to buy.

Our application presented in this paper is the first to combine both Google's Gemini LLM and Apple's ARKit to create an accessible cooking assistant. Compared to previous studies, our application is easier to access due to it being available on the App Store free of charge. We use a more recent CV algorithm from Google's Gemini Flash 1.5 model compared to the YOLOv5 model used by Majil et al.

While this application is beneficial for those who are in need of cooking assistance, the processing speed of the Gemini model, along with the amount of available tokens, serves as a limitation to how responsive the application can be. Additionally, without the use of an AR headset, the application can only go so far when implementing objects into the AR scene. This can limit the potential of having a completely hands-free user experience.

Nevertheless, the application created by this study has the potential to be an accessible cooking option for many, which would result in less hunger and less food wastage. In addition, it would reduce the hassle of cooking for many, making them more eager to cook meals.

## IV. CONCLUSIONS

The research conducted in this paper, along with the application developed, serves as a stepping stone for implementing AR into the cooking experience through accessible mediums such as an iOS device. This application also provides an example usage of the Gemini LLM into accessible cooking technology, as it can be used to scan ingredients, generate recipes, check and provide feedback on cooking steps, and highlight allergens and other potential food hazards.

We tested the impact of our application on the cooking experience of a total of 21 volunteers through 3 separate rounds of surveys. The usability score of the application went from 3.75 to 4.36 over the course of the 3 rounds, indicating that the volunteers were very satisfied by our application in the cooking experience.

By using a similar framework, future studies can use the advances in the depth perception of the cameras in iOS devices to have users interact with AR elements in the AR scene. This would limit interaction with the screen itself, making the cooking experience less messy. Additionally, future works can use even faster and more powerful AI models to have live-labeling of ingredients while still using the knowledge of the Gemini model to generate helpful, creative recipes.

## ACKNOWLEDGEMENTS

I would like to express my gratitude towards Dr. Lina Kim, who provided me with the opportunity to attend the Research Mentorship Program at UC Santa Barbara. Throughout this program, her guidance and teachings on conducting research made my research possible.

Additionally, I am very appreciative of the help provided by my TA, Pratyush Tripathy, who provided me with beneficial advice and feedback throughout my research process.